# Requirements Engineering Practice and Problems in Agile Projects: Results from an International Survey

Stefan Wagner[1], Daniel Méndez Fernández[2],
Michael Felderer[3], Marcos Kalinowski[4]

[1] University of Stuttgart, Institut für Softwaretechnologie, Universitätsstraße 38
D-70569 Stuttgart, Germany
[2] Technical University of Munich, Institut für Informatik, Boltzmannstr. 3
85748 Garching, Germany
[3] University of Innsbruck, Institute of Computer Science, Technikerstr. 21a,
A-6020 Innsbruck, Austria.
[4] Fluminense Federal University, Computing Institute, Av. Milton Tavares de Souza s/n,
Campus Praia Vermelha, 24210-346 Niterói, Brazil.

[1]stefan.wagner@informatik.uni-stuttgart.de, [2]daniel.mendez@tum.de,
[3]michael.felderer@uibk.ac.at, [4]kalinowski@ic.uff.br

**Abstract.** Requirements engineering (RE) is considerably different in agile development than in more traditional development processes. Yet, there is little empirical knowledge on the state of the practice and contemporary problems in agile RE. As part of a bigger survey initiative (Naming the Pain in Requirements Engineering), we build an empirical basis on such aspects of agile RE. Based on the responses of representatives from 92 different organisations, we found that agile RE concentrates on free-text documentation of requirements elicited with a variety of techniques. Often, traces between requirements and code are explicitly managed and also software testing and RE are aligned. Furthermore, continuous improvement of RE is performed due to intrinsic motivation. Important experienced problems include unclear requirements and communication flaws. Overall, we found that most organisations conduct RE in a way we would expect and that agile RE is in several aspects not so different from RE in other development processes.

## 1 Introduction

We have seen a substantial change in the way requirements engineering (RE) is practiced in today's software engineering projects because of the success of agile methods: "No matter the specific method, agile's treatment of requirements is fundamentally different." [11] Furthermore, recent studies indicate that agile practices are frequently adapted to the particularities of their individual environments [6]. However, although we can experience a growth in the body of knowledge about software engineering practices, knowledge on the current state of practice in requirements engineering in general is limited [5]. Moreover, despite the importance of agile practices, little is yet known about how industrial environments conduct RE in an agile setting [3] and what problems they face. Such an understanding would be needed to steer future research in a problem-driven manner.

NaPiRE (Naming the Pain in Requirements Engineering) is an international initiative which tries to fill this gap and to establish a broad survey investigating the status quo of RE in practice together with problems respondents experience in their project environments. In this paper, we investigate RE practice and problems in agile projects based on data from NaPiRE.

## 2  Related Work

We briefly review the existing work on empirical studies on agile requirements engineering before we describe the context and previously published materials.

### 2.1  Empirical Studies on Agile RE

Heikkilä et al. [7] conducted a mapping study on requirements engineering in agile software development in 2015. Hence, it gives a good overview of the topic. They state that "the definition of agile RE is vague." This is reflected in the primary studies that often do not specify the concrete process model used. Most the papers analysed in their mapping study contained some kind of empirical evaluation. We refer to their paper for details. Furthermore, there is a recent systematic literature review by Inayat et al. [8]. They summarise the results of 21 primary studies relating to agile requirements engineering. There is only one paper classified as survey research which in turn conducts interviews.

Cao and Ramesh [3] conducted a qualitative study of 16 software development organisations on their agile RE practices. They identified and rated detailed RE practices. They found, for example, that face-to-face communication, prototyping and reviews and tests are common agile RE practices. To some degree comparable is only the survey by Bustard *et al*. [2]. They investigate the maturity of agile development principles and practices but also touch the topic of requirements. They found that their participants see a process benefit in agile requirements gathering and management. It is also mentioned that while quality requirements were all improved by agile methods, one company stated a "generally weaker treatment of non-functional requirements in an agile approach".

### 2.2  The NaPiRE Initiative

The NaPiRE (Naming the Pain in Requirements Engineering) initiative was started in 2012 in response to the lack of a general empirical basis for RE research. The idea was to establish a broad survey investigating the status quo of RE in practice together with contemporary problems practitioners encounter. This should lead to the identification of interesting further research areas as well as success factors for RE.

We created NaPiRE as a means to collaborate with researchers from all over the world to conduct the survey in different countries. This allows us to investigate RE in various cultural environments and increase the overall sample size. Furthermore, we decided to run the survey every two years so that we can cover slightly different areas over time and have the possibility to observe trends. NaPiRE aims to be open, transparent and anonymous while yielding accurate and valid results.

At present, the NaPiRE initiative has over 50 members from 23 countries mostly from Europe but also North-America, South-America and Asia. There have been two runs of the survey so far. The first was the test run performed only in Germany and in the Netherlands in 2012/13. The second run was performed in 10 countries in 2014/15. All up-to-date information on NaPiRE together with links to instruments used, the data, and all publications is available on the web site http://www.re-survey.org. The first run in Germany together with the overall study design was published in [13] with the detailed data and descriptive analysis available as technical report [12]. It already covered the spectrum of status quo and problems. Overall, we were able to get full responses from 58 companies to test a proposed theory on the status quo in RE. We also made a detailed qualitative analysis of the experienced problems and how they manifest themselves. For the second run, we have published three papers [9, 10, 14] concentrating on specific aspects and the data from only one or two countries and one paper [15] focusing on RE problems, causes and effects based on the complete data set. An analysis of the data with a focus on the state of practice of RE in agile projects has not been published so far.

## 3  Survey Design

This paper uses a part of the overall NaPiRE design: We focus on the descriptive analysis of the state of the practice and potential problems in agile requirements engineering. For that, we analyse the data from the second NaPiRE run conducted in 2014/15. In the following, we detail the information on the study design relevant to the analysis presented in this paper.

### 3.1  Research Questions

We aim at understanding the state of practice of requirements engineering in agile projects. This cannot be exhaustive as there are too many aspects potentially relevant to agile requirements engineering. Our objective is to be generic to be able to apply the same instrument to non-agile projects. To this end, we formulate the following four research questions, shown in Table 1, to steer the design of our study.

**Table 1.** Research Questions.

| | |
|---|---|
| RQ 1 | How are requirements elicited and documented? |
| RQ 2 | How are requirements changed and aligned with tests? |
| RQ 3 | Why and how is RE improved? |
| RQ 4 | What are common problems in agile RE? |

The first question aims to capture the most basic activities in RE: elicitation and documentation. Yet, a key principle in agile development is that requirements are not stable. Hence, we want to understand how agile projects deal in particular with changing requirements. A further key principle in agile development is the continuous improvement of the development process itself. This should also hold for the RE process. Therefore, we are interested in whether agile projects perform continuous improvement and what is their motivation. Finally, after gathering an understanding about the state of the practice, we want to understand how important various potential problems for RE are in agile projects and what are their causes and effects.

### 3.2 Instrument

The instrument used in NaPiRE constitutes in total 35 questions used to collect data on topics including the demographics, how practitioners elicit and document requirements and finally what problems practitioners experience in their RE. In this study, we focus on the status quo using the demographics only as context and to select the companies working in an agile manner. We will also discuss the main problems as rated by these companies, but we will not go into a more detailed problem analysis. Table 2 summarises the excerpt of our questionnaire in scope of this study.

**Table 2.** Questions (simplified and condensed excerpt).

| Parts | No. | Question | Type |
| --- | --- | --- | --- |
| Demographics | Q 1 | What is the size of your company? | Closed (SC) |
| | … | … | … |
| | Q 8 | Which process model do you follow (or a variation of it)? | Closed (MC) |
| Status Quo | Q 9 | How do you elicit requirements? | Closed(MC) |
| | Q 10 | How do you document functional requirements? | Closed(SC) |
| | Q 11 | How do you document non-functional requirements? | Closed(SC) |
| | Q 12 | How do you deal with changing requirements after the initial release? | Closed(SC) |
| | Q 13 | Which traces do you explicitly manage? | Closed(MC) |
| | Q 14 | How do you analyse the effect of changes to requirements? | Closed(MC) |
| | Q 15 | How do you align the software test with the requirements? | Closed(MC) |
| | … | … | … |
| | Q 23 | Is your RE continuously improved? | Closed(SC) |
| | Q 24 | Why do you continuously improve your RE? | Closed(MC) |
| | … | … | … |
| Problems | Q 28 | Considering your personal experiences, how do the following (more general) problems in requirements engineering apply to your projects? | Likert |
| | … | … | … |

For these areas, we only use closed questions. The answers can be mutually exclusive single choice or multiple choice answers. Most of the closed multiple choice questions include a free text option, e.g. "other", so that the respondents can express company-specific deviations. We furthermore use Likert-type scales on an ordinal scale of 5 and define for each a maximum value (e.g., "agree", or "very important"), a minimum value (e.g., "disagree", or "very unimportant"), and the middle ("neutral"). These are used to answer the last question on the problems where we let the respondents rate the extent to which a given set of typical RE problems apply to their agile project environments.

### 3.3 Data Collection

The survey is conducted by invitation only to have a better control over the distribution of the survey among specific companies and also to control the response rate. The responses were, however, anonymous to allow our respondents to freely share their experiences made within their respective company. For each company, we invited one respondent as a representative of the company. In case of large companies involving several autonomous business units working each in a different industrial sector, we selected a representative for a unit. For the data collection, each country representative defined an invitation list including contacts from different companies and initiated the data collection independently as an own survey (sub-)project. All surveys relied on the same survey tool hosted and administrated by the authors.

We conducted the survey in North America (Canada, USA), South America (Brazil), Central Europe (Austria, Germany, Ireland) and Northern Europe (Estonia, Finland, Norway, Sweden). The data collection phases in each country and response rates are shown in Table 3.

**Table 3.** Data collection phase (overview).

| Area | Country | Data Collection Phase | Response Rate |
|---|---|---|---|
| Central Europe | Austria | 2014-05-07 to 2014-09-15 | 72.0 % |
| | Germany | 2014-05-07 to 2014-08-18 | 36.8 % |
| | Ireland | 2014-05-07 to 2014-12-31 | 39.7 % |
| North America | Canada | 2014-05-07 to 2015-08-15 | 75.0 % |
| | USA | 2014-05-07 to 2015-05-01 | 60.0 % |
| Northern Europe | Estonia | 2014-05-07 to 2014-10-31 | 89.0 % |
| | Finland | 2015-06-01 to 2015-08-28 | 83.0 % |
| | Norway | 2014-05-07 to 2014-09-15 | 59.0 % |
| | Sweden | 2014-05-07 to 2014-09-15 | 34.0 % |
| South America | Brazil | 2014-12-09 to 2015-03-31 | 63.0 % |

### 3.4 Data Analysis and Validity Procedures

In the subset of NaPiRE that we will discuss in this paper, we conduct two types of analysis: The first analysis is frequency counting for questions in which the respondents choose one or more options. This is, for example, the case when they should choose which requirements elicitation techniques they use. We extract the counts using an R script which also creates bar charts from it.

The second analysis is necessary for the question about contemporary problems. We analyse the Likert type data by transforming the answers to numbers from 1 to 5. Then, we calculate and report the median and the median absolute deviation (MAD) for each problem also using an R script. We refrain from a detailed qualitative analysis and coding of the free-text answers, because this is out of scope of this paper.

Yet, we use them to substantiate the discussion and interpretation of the ranking of importance of the problems. The overall NaPiRE endeavor includes several procedures for checking validity, i.e., concerning the data collection and analysis phases, as described in detail in our previously published material [13].

## 4 Results

In the following, we summarise our results structured according to the research questions and beginning with an overview of the study population.

### 4.1 Study Population

Overall, we received 354 answers to the second NaPiRE run in 2014/15 out of which 228 completed the questionnaire. Out of these, we selected the 92 organisations that answered "Scrum" and/or "XP" as their development process model, but not "Waterfall", "V-Model XT" or "Rational Unified Process". Hence, the following results represent the situation of 92 different companies or business units (in case of large companies).

To better illustrate the study population, we grouped organisations into small, medium, and large ones. For this grouping, we relied on the number of employees. Organisations with up to 50 employees were considered small, with 51 to 250 medium, and organisations with more than 250 were considered large. Table 4 summarises the distribution of the responses according to the different company sizes and the areas where they are situated.

Table 4. Responding organisations by size and region.

| Size | Central Europe | North America | Northern Europe | South America | Total |
|---|---|---|---|---|---|
| Small | 6 | 4 | 6 | 14 | 30 |
| Medium | 4 | 0 | 8 | 10 | 22 |
| Large | 12 | 8 | 11 | 8 | 39 |
| Unknown | 0 | 0 | 1 | 0 | 1 |
| | 22 | 12 | 26 | 32 | 92 |

### 4.2 Elicitation and Documentation (RQ 1)

We start answering RQ 1 by looking at how agile projects elicit requirements. We used the elicitation technique classification as provided in the SWEBOK (www.swebok.org). How often these elicitation techniques have been selected by our respondents is shown in Fig. 1. The most frequently used techniques are interviews, prototyping, and facilitated meetings. Scenarios are employed by about half of the respondents, observations by less than a third.

We believe these answers fit very well to the expectations on agile projects. Roles like a product owner in Scrum would use interviews to understand the overall product requirements while the further elicitation is done in workshops with stakeholders and the sprint planning. Prototyping is usually not an explicit part of agile methods but building minimal viable products could be seen as a form of prototyping. Furthermore, paper prototypes or wire frames of user interfaces can also be useful in agile projects. Observations are not frequently used. Maybe there is potential to explore this kind of elicitation in more detail. But it might also be caused by a lack of possibilities for the developers.

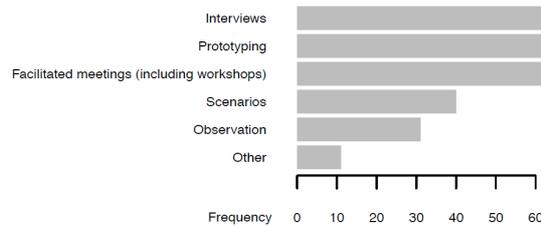

**Fig. 1.** How do you elicit requirements?

Next, we asked about the documentation of the most frequent type of requirements: functional requirements. The respondents could choose multiple items from different description techniques, namely (structured requirements lists, domain/business process models, goal models, data models, and use case models) as well as the degree of formality (free form, textual, textual with constraints, and semi-formal or formal).

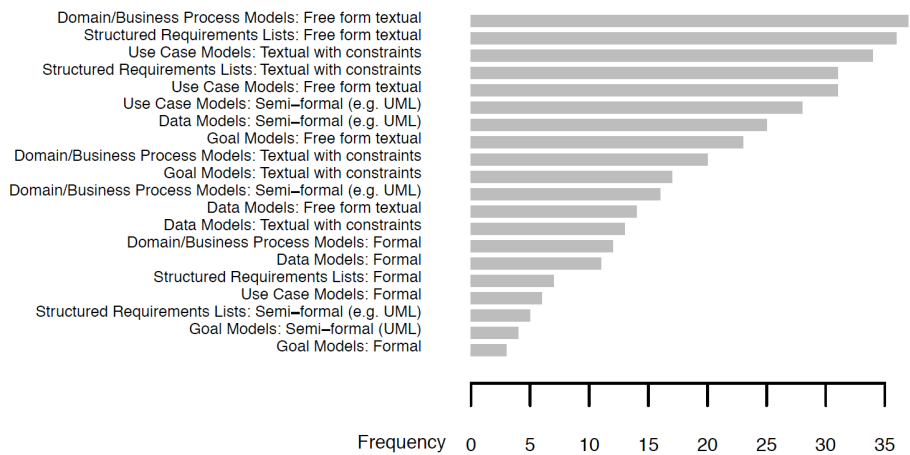

**Fig. 2.** How do you document functional requirements?

As shown in Fig. 2, the three most frequent ways to document requirements are as free-form textual domain/business process models, free-form textual structured requirements lists and use case models as text with constraints. But also structured requirements lists as texts with constraints and free-form textual use case models are used by almost a third of the respondents. Data models are almost only used in a semi-formal notation such as the UML. Goal models are rarely used overall. Formal notations for requirements are also rarely used.

This again fits to the expectation of common agile methods: requirements are usually written down as text either in a free form or with some constraints (such as the role/feature/reason schema for user stories). Only data models are documented with a class diagrams or a variation of them. More semi-formal and formal documentation methods are probably too heavy-weight or unnecessary in the presence of automated tests for requirements. Especially the role of automated tests would be interesting to follow-up in further studies.

Finally, we briefly touched also the topic of non-functional requirements (such as security or performance requirements). We found that most respondents document non-functional requirements with text. About half of those document non-functional requirements either in a quantified manner, e.g., by defining concrete measurements, or in a non-quantified manner, e.g., by linking to external reference models or style guides.

### 4.3 Changing Requirements (RQ 2)

In RQ 2, we are interested in how agile projects document changes in requirements. First, we asked how the respondents deal with changing requirements after the initial release. The answers are shown in Fig. 3. As to be expected, the overwhelming majority updates the product backlog when requirements change. Yet, 16 % only work with change requests and 15 % even have a requirements specification they regularly change. Overall, the product backlog seems to be the common way to work with changing requirements in agile projects, but it is not always clear how it works together with change requests.

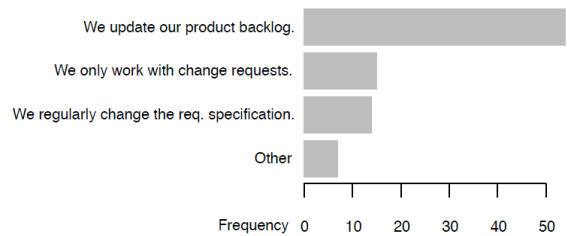

**Fig. 3.** How do you deal with changing requirements after the initial release?

Next, we were interested in how the respondents analyse the effect of changes to requirements. As shown in Fig. 4, most respondents do impact analysis between requirements. More than a third analyse the impact of requirement changes on the code. A fifth do no analysis of the effect of changes to the requirements. Answers for *Other* included "test-driven analysis for TDD projects", "rerun test suites", "we discuss with users and decide the best approach" and "team-based discussion before change". Therefore, besides looking at requirements and code, the test suites and direct discussions with stakeholders seem important for impact analyses in agile projects.

A help for impact analysis are traces between requirements and code or between requirements and design documents. Concerning this question, more than half of the respondents answered that they explicitly manage traces between the requirements and the code. A third manages explicitly the traces between requirements and design documents. More than a fifth of the respondents do not explicitly manage traces at all.

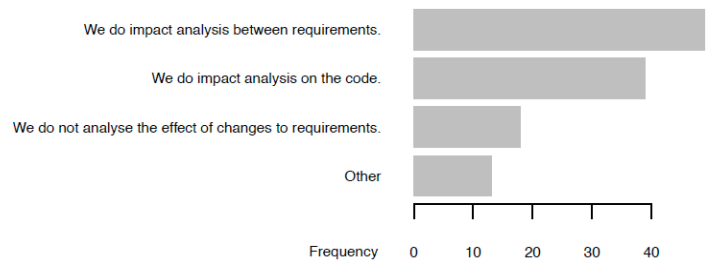

**Fig. 4.** How do you analyse the effect of changes to requirements?

Finally, we had a question relating requirements and tests. We asked how the respondents align the tests with the requirements. As shown in Fig. 5, in agile RE, it is common to define acceptance criteria. This is what we would expect because specific test-driven practices which have become popular in and through agile methodologies like test-driven development [1] and behaviour-driven development [4] as well as the common user story practices demand to make acceptance criteria explicit. Furthermore, also coverage of requirements by tests is considered in a remarkable number of agile projects. Also in this case, test-driven practices linked to agile methodologies may be a trigger for that. In about half of the projects of the respondents, the testers participate in requirements reviews. This also means that half of the projects do have requirements reviews, which we would not expect from all agile projects as it is not demanded in common agile development processes. Finally, The derivation of tests from system models is only rarely done.

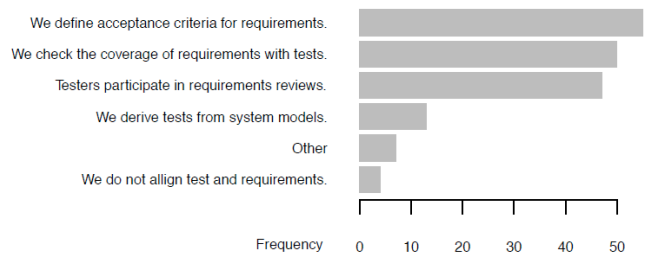

**Fig. 5.** How do you align the software test with the requirements?

### 4.4 RE Improvement (RQ 3)

Also, and maybe in particular, requirements engineering processes need to be improved. In an agile context, we would expect this improvement to be done continuously. We asked whether the organisations improve their RE continuously and who is responsible for this improvement. The results in Fig. 6 show that in more than half of the responding organisations, the RE is continuously improved and this improvement is under sovereignty of the project team. This is in tune with our expectations because of the deeply entrenched idea to regularly work on the development processes with, for example, retrospectives in Scrum.

Yet, also almost a third of the respondents have an own business unit or role responsible for the continuous improvement. Only few respondents use external consultants for that. Still, about 14 % of the respondents do not continuously improve their RE.

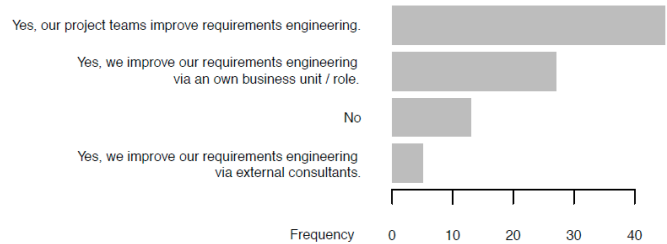

**Fig. 6.** Is your requirements engineering continuously improved?

At this point, we wanted to dig a bit deeper and understand the reasoning behind doing continuous improvement. As shown in Fig. 7, most of the respondents who do it because it helps them to determine their individual strengths and weaknesses and to act accordingly. Hence, the motivation is mostly intrinsic. Only a quarter or below give extrinsic reasons such as the expectation of the customer, certifications or regulations.

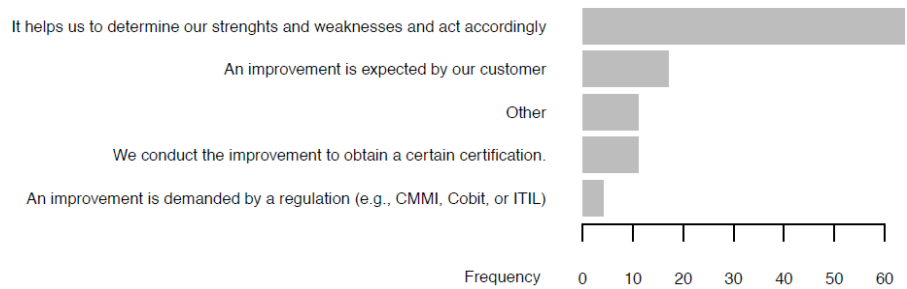

**Fig. 7.** Is your requirements engineering continuously improved?

Hence, continuous improvement in general as well as in RE is widespread in agile projects in practice. The motivation is intrinsic based on a perceived improved efficiency and because it is postulated by agile process models.

### 4.5 Problems in Agile RE (RQ 4)

Finally, after getting an overview of the current state of practice (RQ1–3), we wanted to know what common problems the respondents experience in their respective project environments. To this end, we presented a list of common RE problems and asked the respondents whether they agree that these problems occur in their setting. Table 5 summarises the problems, ordered from top to bottom according to the agreement by the respondents.

The problems ranked as low where not surprising to us considering the agile RE setting. For instance, volatility in the customer's business domain seems not to be a critical problem. Indeed, changes in processes or requirements are what agile processes are designed for and this shows in the little relevance of this problem in practice.

Table 5. Considering your personal experiences, how do the following problems in requirements engineering apply to your projects? (from 1: I disagree to 5: I agree).

| Problem | Median | MAD |
|---|---|---|
| Underspecified reqs. that are too abstract and allow for various interpretations | 4 | 1 |
| Unclear / unmeasurable non-functional requirements | 4 | 1 |
| Communication flaws within the project development team | 4 | 1 |
| Communication flaws between developers and the customer | 4 | 1 |
| Moving targets (changing goals, business processes and / or requirements) | 4 | 1 |
| Incomplete and / or hidden requirements | 4 | 1 |
| Implicit requirements not made explicit | 4 | 1 |
| Stakeholders with difficulties in separating reqs from previously known solutions | 4 | 1 |
| Inconsistent requirements | 4 | 1 |
| Insufficient support by project lead | 3 | 1 |
| Insufficient support by customer | 3 | 1 |
| Missing traceability | 3 | 1 |
| "Gold plating" (implementation of features without corresponding requirements) | 3 | 1 |
| Weak access to customer needs and / or (internal) business information | 3 | 1 |
| Weak knowledge of customer's application domain | 3 | 1 |
| Weak relationship between customer and project lead | 3 | 1 |
| Time boxing / Not enough time in general | 3 | 1 |
| Discrepancy between high innovation and need for formal acceptance of requirements | 3 | 1 |
| Volatile customer's business domain regarding | 3 | 1 |
| Terminological problems | 3 | 1 |
| Unclear responsibilities | 3 | 1 |
| Technically unfeasible requirements | 2 | 1 |

Similarly, unclear responsibilities are rarely experienced as a problem. The clear roles in agile processes seem to provide a good understanding here. Respondents who experienced this problem informed problems on the developer side or customer side to really understand and live up to their corresponding roles. Hence, the roles in agile RE seem to support clear responsibilities but they need to be clearly understood.

Some of the top ranked problems in turn can be argued based on the current state resulting from the agile process models used, such as unclear / unmeasurable non-functional requirements or underspecified requirements. The latter is caused by general problems in the capabilities of the involved people either on the development team ("Ability to write requirements and analyse customer needs") or on the customer side ("Customer clueless about functions of the system"). But also communication between the developers and the customers seem to cause these problems ("Developer may do their own wrong interpretation"). Furthermore, also in agile projects, it seems to be problematic to rush too quickly through defining what needs to be done ("Not enough time spent defining to the level of detail required").

For unclear or unmeasurable non-functional requirements, the cause seems to be mostly the reliance on experience ("Boils down to experience, on both ends. Non-functional requirements are easy to miss.") and unsuccessful communication ("think we talk about the same thing but not"). Agile RE might be able to provide more structure and terminology in this area to avoid consequences such as "Lots of surprises in deployment" and "unhappy end-users".

Some of the top-ranked problems, however, are at the same time surprising to us given that those problems also form a natural condition for agile projects. We refer in particular to moving targets and incomplete requirements which are stated as problems and which should motivate the use of agile practices. Even more, communication flaws within the project development team as well as communication flaws between developers and the customer are stated under the top problems.

The moving target problem is caused by "Changing priorities", "Changes and instability in the customer organization are not isolated from our process. Their problems leak through." and "Mostly the reason is that the business is constantly learning at the same time or changes in management." This leads to "already specified requirements may become obsolete". Hence, fixing requirements during a sprint as, for example, emphasised in Scrum seems important to address these causes. Yet, overall the effects can be small: "If parties are on board that things are changing then the project won't have problems in term of budget, timeline etc. because everybody knows these are flexible as long as targets are moving. It will cause stress for dev team though."

In many agile RE approaches, requirements are not meant to be complete but a cause for discussions with the customer. Hence, incomplete requirements are to be expected. When does this become a problem? It is a problem if the effect is "Rework or delivery that does not fully meet the customer's need." or "customer dissatisfaction (delivery that does not meet customer expectations)". It is caused by "Hidden requirements that are obvious to the customer" and inexperience of the product owner and customer ("Lack of experience of the Product Owner; Lack of clarity/understanding of the client"). Hence, the role of the on-site customer or product owner is a central one that needs to be filled with a person being able to understand the customers and elicit all important requirements.

The communication flaws seem to be mostly caused by missing time ("Also related to an attempt to gain time in developing.", "high need for meetings and documentation versus time") and more general communication problems ("Lack of open dialogue on the team.", "Our developers don't know the flows to generate questions to other teams."). These communication problems can lead to unnecessary work ("we waste time trying to develop new features that were developed by other teams previously") and generating unnecessary risks ("Unsolved problems due to the lack of dialogue between people.").

We interpret this as that the prerequisite on which agile RE relies, i.e. human-intensive continuous exchange, can quickly manifest itself as a critical problem. That is, agile RE does not necessarily solve all problems plan-driven process models often have, but they become explicit once key prerequisites for successful RE are not met: human-intensive exchange and collaboration. Yet, it would be interesting to understand in more detail the causes and effects of these problems in agile projects.

## 5   Conclusions and Future Work

In this paper, we reported on the results from an analysis of the current state of practice and potential problems in agile requirements engineering. Our analysis is based on data gathered from a globally distributed family of practitioner surveys (called NaPiRE). We shed some light on how requirements are elicited and documented, how our respondents deal with changing requirements, why and how RE is improved, and on common RE problems.

Overall, we found that most of the responding organisations conduct RE in a way we would expect in agile projects. The documentation of requirements is dominated by free-text documents with some constraints. The backlog is the central means to deal with changing requirements. Code and requirements are explicitly linked and RE is continuously improved because of an intrinsic motivation.

Yet, for all these aspects there is also a considerable number of projects claiming to follow Scrum or Extreme Programming and not working in that way. To some degree, this supports or findings from [6] that many companies claiming that they do Scrum actually deviate heavily from it. Therefore, in future surveys, we will need to differentiate in more detail.

In terms of the current state of practice concerning elicitation, documentation, changes and improvement, agile RE is not so different from classical RE after all. Our concluding analysis of contemporary problems in RE revealed that some of the problems often seen to come along the use of plan-driven process models are not seen as critical anymore. Others, however, which are often seen as a motivation for agile RE, e.g., moving targets, can still become dominant. We will have to dig deeper into specifics of agile RE to better understand what agile RE practices are related to which problems or their mitigation.

Although our analysis is based on a broad family of surveys, we are aware that our study has limitations. First, our results emerge from a reasonable but still limited sample with a limited context model. We therefore cannot make concrete statements about how generalisable the results eventually are, let alone because we still are not able to estimate the representativeness of our population. Therefore, we need to follow our design of a family of surveys and further steer our continuous replications while capturing the context more precisely.

Also, inherent to survey research is that surveys can only reveal stakeholders' perceptions on current practices rather than empirically backed-up knowledge about those practices. Although we were interested in revealing those perceptions, the answers given by our respondents might still be biased. We mitigated this threat by conducting the survey anonymously, but need to apply further empirical methods in the future to further explore the field based on project data.

Finally, NaPiRE was not intentionally designed to explore RE in agile contexts. This has two implications. First, the selection of our sample was based on the self-assessment of the respondents based on a predefined list of options in the process models. Second, although we could analyse different variations in the status quo of how our respondents do their RE, our instrument does not yet capture the particularities of agile practices (e.g. considering agile artefacts such as user stories). A richer investigation of facets important to agile RE forms part of future work where we redesign the instrument to give more attention to agile practices.

**Acknowledgments.** We are grateful to the whole NaPiRE team as well as all participating respondents.